\documentclass[conference]{IEEEtran}
\IEEEoverridecommandlockouts

\usepackage{cite}
\usepackage{amsmath,amssymb,amsfonts}
\usepackage{graphicx}
\usepackage{textcomp}
\usepackage{xspace}
\usepackage{trimclip} 
\usepackage{trimclip} 
\usepackage{mathtools}
\usepackage{algorithm}
\usepackage{algpseudocode}
\usepackage{subcaption}
\usepackage{multirow}
\usepackage{booktabs}
\usepackage{arydshln}

\usepackage{dsfont}
\usepackage{upgreek}
\usepackage{bbm}			%
\usepackage{bbold}		
\usepackage{xcolor}		
\usepackage{hyperref}
\hypersetup{
 colorlinks=true,breaklinks,
 linktoc=section,
 linkcolor=red,
 citecolor=blue,
 urlcolor=blue,
 }

\usepackage{amsthm}
\usepackage{thmtools}

\usepackage[noabbrev]{cleveref}

\def\BibTeX{{\rm B\kern-.05em{\sc i\kern-.025em b}\kern-.08em
 T\kern-.1667em\lower.7ex\hbox{E}\kern-.125emX}}

\newtheorem{proposition}{Proposition}
\newtheorem{definition}{Definition}
\newtheorem{remark}{Remark}

\newcommand{\Sec}[1]		{Sec.\,\ref{#1}}
\newcommand{\Fig}[1]		{Fig.\,\ref{#1}}

\newcommand{\Eq}[1]			{Eq.\,\ref{#1}}
\newcommand{\Tab}[1]		{Tab.\,\ref{#1}}
\newcommand{\Alg}[1]		{Alg.\,\ref{#1}}

\newcommand{\Proposition}[1]{Proposition\,\ref{#1}}

\newcommand{\Definition}[1]{Definition\,\ref{#1}}

\newcommand{\Remark}[1]{Remark\,\ref{#1}}
\newcommand{\ie} 			{i.e.\xspace}
\newcommand{\eg} 			{e.g.\xspace}

\newcommand{\iid} 		{i.i.d.\xspace}
\newcommand{\vs} 		 {vs.\xspace}
\newcommand{\st} 			{\mbox{s.t.}\xspace}

\newcommand{\ind} {\mathds{1}}%

\newcommand{\Exp}[1] {\mathbbm{E}[#1]}
\newcommand{\real} {\mathbbm{R}}
\newcommand{\Prob} {\mathbbm{P}}

\newcommand{\mydots} 	{...}

\newcommand{\INTerval}[2] 	{\{#1,\mydots, #2\}} %

\newcommand{\subtitle}[1] {{\emph{\textbf{#1}}}}

\DeclareMathOperator*{\argmax}{\arg\!\max}
\DeclareMathOperator*{\argmin}{\arg\!\min}

\usepackage[normalem]{ulem} %

\newcommand{\myEndBlock} {{\footnotesize$\triangle$}}

\newcommand{\myEndBlockProof} {{\footnotesize$\square$}}

\newcounter{marginNoteCounter}

\newcommand{\International} {Intern.\xspace}
\newcommand{\Conference} {Conf.\xspace}
\newcommand{\Symposium} {Symp.\xspace}
\newcommand{\Transactions} {Trans.\xspace}
\newcommand{\Proceedings} {}%
\newcommand{\Journal} {Journal\xspace}

\usepackage{balance}

\newcommand{\sample} 		 {\sample\xspace}

\renewcommand{\sample} 	 {sample\xspace}

\newcommand{\aj}		 {\tilde{A}_{j}}

\newcommand{\Xbold}	 {\mathbf{X}} %
\newcommand{\val}			 {S} %
\newcommand{\Sbold}	 {\mathbf{\val}} %

\renewcommand{\S}[1]				{ \val_{#1} }

\newcommand{\Mboldround}[1]		{\mathbf{M}_{#1}}

\newcommand{\Xrefbold}[1]		 { \mathbf{S}^{\mathrm{upd}} } 
\newcommand{\Xref}[2]				{S_{(#1)}^{\mathrm{upd}} } 
\newcommand{\symb}[1]		{\dot}
\newcommand{\Xobold}		 {\symb{\mathbf{X}}}%
\newcommand{\Abold}		 { \mathbf{A} } %

\newcommand{\rsymb} 		 	 {\phi} %

\newcommand{\thresh} 			{S^{\mathrm{acc}} }%

\newcommand{\mydef} 			{:=}%

\newcommand{\algo}[1]		{{\textsc{#1}}}

\newcommand{\MEAN}			 {\algo{mean}\xspace}	
\newcommand{\MEDIAN}			 {\algo{median}\xspace}
\newcommand{\OPTIMISTIC}			{\algo{optimistic}\xspace}
\newcommand{\VIRTUAL}			{\algo{virtual}\xspace}
\newcommand{\VIRTUALp}			{\algo{virtual+}\xspace}
\newcommand{\SREF}			{\algo{single-ref}\xspace}
\newcommand{\KLEINBERG}			{\algo{kleinberg}\xspace}

\newcommand{\DynSIMPLEK}			{\algo{dyn-simplek}\xspace}
\newcommand{\SIMPLEK}			{\algo{simplek}\xspace}
\newcommand{\GREEDY}			{\algo{greedy-offline}\xspace}
\newcommand{\frate}			{\rho_f}

\newcommand{\FRMfull}			 {Failure Rate Minimization\xspace} %
\newcommand{\FRM}			 			{\algo{frm}\xspace}
\newcommand{\FRMbasic}			{\algo{frm}-{\small \algo{basic}}\xspace}
\newcommand{\FRMnooverflow}	{\algo{frm}-{\small \algo{no}}-{\small \algo{overflow}}\xspace}

\renewcommand{\symb}[1]			{\dot{#1}}
\newcommand{\mindist}		{d}%
\newcommand{\xo}[1]			{\symb{x}_{#1}}

\newcommand{\btot}		{b} %
\renewcommand{\Abold}[1]		{\mathbf{A}_{#1}}

\newcommand{\SPfull}		{Secretary Problem\xspace}

\newcommand{\instances} {items\xspace}

\newcommand{\m} 	 {m}
\renewcommand{\aj} 			{\sum_{i=c+1}^{j} A_i}

\renewcommand{\thresh} 			{\tau}
\newcommand{\switch}			{j^*} %
\renewcommand{\Xref}[2] 			{ S^\textup{ord}_{#1,#2} }
\renewcommand{\Xrefbold}[1] 			{ \mathbf{S}^\textup{ord}_{#1} }

\newcommand{\Dist} 				{\operatorname{dist}}
\renewcommand{\mindist}			 {D}	
\newcommand{\dec}			 {q}	
\newcommand{\sep}			{\hspace{2mm}}
\newcommand{\Synth}			{Synth}	

\newcommand{\MaxMin}			{max-min\xspace}
\newcommand{\MaxSum}			{max-sum\xspace}

\newcommand{\streambased} {streaming\xspace}

\newcounter{phase}[algorithm]
\newlength{\phaserulewidth}
\newcommand{\setphaserulewidth}{\setlength{\phaserulewidth}}
\newcommand{\phase}[1]{%
 \vspace{-1.45ex}
 \Statex\leavevmode\llap{\rule{\dimexpr\labelwidth+\labelsep}{\phaserulewidth}}\rule{\linewidth}{\phaserulewidth}
 \Statex\strut\refstepcounter{phase}\!\!\!\!\!\!\!\!\raisebox{0.25em}{{\scriptsize$_\blacksquare$}}\ \textit{\textbf{#1}}%
 \vspace{-1.25ex}\Statex\leavevmode\llap{\rule{\dimexpr\labelwidth+\labelsep}{\phaserulewidth}}\rule{\linewidth}{\phaserulewidth}}
\makeatother
\setphaserulewidth{.3pt}
\begin{document}

\title{Efficient Stream-based Max-Min Diversification \\with Minimal Failure Rate\\

\author{%
\IEEEauthorblockN{Argyris Kalogeratos\qquad Yutai Nazir Zhao\qquad Mathilde Fekom}
\IEEEauthorblockA{\textit{Centre Borelli, ENS Paris-Saclay, 91190 Gif-sur-Yvette, France}}
}
\thanks{$\square$~The authors acknowledge the support from the Industrial Data Analytics and Machine Learning Chair hosted at ENS Paris-Saclay. Correspondence to: \\
argyris.kalogeratos@ens-paris-saclay.fr.}
}

\maketitle

\begin{abstract} 
The \textit{\streambased \MaxMin diversification problem} concerns the selection of a limited and diverse sample of items out of a data stream of known finite length. %
The %
objective to be maximized is the minimum distance among any pair of selected items.
We consider the \emph{irrevocable-choice sampling}, where decisions need to be immediate and irrevocable while processing the items of the stream, which is a setting little studied in the literature. %
Standard algorithmic approaches for sequential selection disregard \textit{selection failures}, which is when the last items of the stream are picked by default, to prevent delivering an incomplete selection set. %
This defect can be catastrophic for the \MaxMin diversification objective. %
The proposed \emph{\FRMfull} (\FRM) is a rank-based algorithm that selects a set of diverse \instances and, in addition, reduces significantly the probability of having failures. %
We demonstrate with simulations \FRM's %
performance comparing with existing selection strategies.
\end{abstract}

\begin{IEEEkeywords}
Data streams, \MaxMin diversification, online selection, robotic sampling, sequential selection 
\end{IEEEkeywords}

\section{Introduction}
In certain applications, data are presented to the system as a \emph{stream} of sequentially arriving \emph{items} \cite{Vitter85, Muthukrishnan05}; for example, when new items are becoming available in a large information platform, or when observations are made by an autonomous robot while exploring an environment \cite{OODS2009,OnlineNavigation2010}.
Long or high-frequency data streams are difficult to handle by machines of limited computational %
and storage capacity. %
Aside the need for %
immediate decisions, as streams cannot be stored, there is often the requirement for outputting results online while processing the stream \cite{IncrDiv2011}. %
Most typically, %
this includes respecting memory limitations and time constraints, the dynamic adaptation of learning algorithms, or dealing with non-stationarity where the properties of the incoming data change over time. %
\emph{Irrevocable-choice sampling} \cite{Zhu16, irrevocable-sampling-periodic-data-2018} emerges in the context of stream processing. It refers to the online selection of a limited-sized sample of exactly $b$ items from an input stream of length $N$, with decisions that need to be \emph{immediate} and \emph{irrevocable}. The \emph{budget} $b$ is small and fixed in advance. %
The first requirement is standard when rejected items cannot be recalled \cite{IncrDiv2011} after the examination (and rejection). The second one emerges when discarding already selected items is not an option. Note also that, when operating under such constraints there is a risk of \emph{selection failure}, which occurs when the last incoming
items in the sequence are selected by default
in order to prevent delivering an incomplete selection.

Irrevocability can be crucial in several settings, notably in frugal systems where processing each selected item can be costly energy- and time-wise, resources that are extremely limited in that setting. 
Moreover, an information system may require each selected item to be outputted immediately %
further up to a master process, which in turn needs to perform subsequent tasks. In that case, revoking and revising a selection can be very complicated or infeasible. %
Revoking selected items can also be difficult when deploying autonomous robotic agents that %
pick real items, such as materials or living samples, while exploring a natural environment. The discharge of such an item may alter or disturb the local environment where the agent is located, which is different from where that item was collected. For instance, \cite{Zhu16} investigated an application where the stream represents insects or fish types that are sequentially discovered by an autonomous robotic device following a predefined exploration path, and decides immediately and irrevocably whether to collect or not each examined item. Irrevocable marine robotic sampling was also investigated in \cite{%
OnlineNavigation2010, irrevocable-sampling-periodic-data-2018}, where the latter focuses on periodic streams.%
In this paper, we focus on an online version of the \emph{$b$-diversification} problem \cite{Chandrasekaran81, Akagi18} (a.k.a. \emph{result diversification} or \emph{$b$-dispersion} problem). In the batch version, a small and as diverse as possible subset needs to be selected out of a large pool of available items. %
The output set shall contain various different elements, which is particularly useful when returning to a user the results of his search queries, or recommendations \cite{Ziegler05, Vee08, Yu09, Vargas11}. 
The diversity, often modeled by a distance measure, may represent \emph{dissimilarity}, \emph{novelty}, or \emph{semantic coverage} of the compared items, and there are several measures to quantify it for a set \cite{Drosou12, Drosou17, core-set-coverage02014}. 
For instance, a well-studied objective is the \emph{\MaxSum diversification} \cite{Borodin12} that maximizes the sum of the utility of the selected items, whereas a hardly studied objective is the \emph{\MaxMin diversification} \cite{Zhu16} that maximizes the minimum distance between any two selected items. %
This batch diversification problem has is %
NP-hard, however efficient approximations do exist \cite{Erkut94}. %

Focusing on the online case, we present the \emph{\FRMfull} (\FRM), a rank-based and purely online selection algorithm to address the \streambased \MaxMin diversification problem, while reducing the \emph{failure rate}. %
A failure can dramatically reduce the quality of the final selection set, especially when assessed by the \MaxMin objective function, as even one failure could deteriorate the quality of the entire set. \FRM is put in comparison with existing algorithms where it shows its superior trade-off between time complexity and performance.

\section{Streaming \MaxMin diversification}\label{sec:diversification_problem} %
\subsection{Notations and problem definition}\label{sec:problem_def}
Let $x_j \in \real^{d}$ be the $j$-th arriving item in a finite stream of $N\in \mathbbm{N}^*$ independent and identically distributed (\iid) items; let $\Xbold \in \real^{d\times N}$ be a matrix representing the full stream. %
The sequential selection problem aims to select online a small subset $\Xobold_b \subseteq \Xbold$ of $b \in \mathbbm{N}^*$ items, typically $\btot \ll N$, that satisfy a given objective.
At step $j$ of the process, the selection set has $k$ items: $\Xobold_k = \{\xo{1},\mydots,\xo{k}\} \subseteq \Xbold$, where $k\le j$ and $k\le b$. Let $\Mboldround{k} = \{\m_1,\mydots,\m_k\}$ be those items' positions in the sequence, \ie $\xo{l}=x_{\m_{l}},\,\forall l \le k$. The process is subject to three constraints: (C1)~\emph{immediate decisions} forbid rejected items $x_l$, $l < j$, to be recalled; (C2)~\emph{irrevocable decisions} allow the selection set to only grow or remain unchanged, \ie $\Xobold_{k+1} \supseteq \Xobold_k$; (C3)~\emph{selection completeness} requires selecting exactly $b$ items.
At the end of the process, $\Xobold \mydef \Xobold_b$, $\Mboldround{} \mydef \Mboldround{b}$, and the decisions taken are denoted by $\Abold{} : A_{j} = 1$, if $j \in \Mboldround{}$, otherwise $A_{j} = 0$, %
$1 \leq j \leq N$.

In our problem of interest, initially only the following parameters are known: the budget size $b$, the stream length $N$, and a pairwise distance measure $\Dist(\cdot,\cdot) \in \real_+$. The objective, formally expressed in \Definition{def:streaming-max-min-div}, is that the minimum distance between any two items of the final selection set is maximized. 
Let the minimum distance between pairs of items of two %
input sets $\mathbf{U}$ and $\mathbf{V}$ be defined as:
\begin{equation}\label{eq:set_dist}
\mindist(\mathbf{U},\mathbf{V}) = \min_{x_i \in \mathbf{U},\, x_j\in \mathbf{V}}
\Dist(x_i,x_j).
\end{equation}
For a single input set, %
the above reduces to: %
\begin{equation}
\mindist(\mathbf{V}) %
= \min_{x_i,x_j\in \mathbf{V},\, i\neq j} \Dist(x_i,x_j).
\end{equation}
In the rest, we refer to the particular case where the set $\mathbf{U} = \{x\}$ has always cardinality one, thus we simplify the associated notation of \Eq{eq:set_dist} (and similarly other deriving ones) by writing $\mindist(x, \mathbf{V}) \equiv \mindist(\{x\}, \mathbf{V})$.
\begin{definition}\label{def:streaming-max-min-div}%
\textbf{Streaming \MaxMin $b$-diversification.}~The objective is to find the $b$-sized subset $\Xobold_b^* \subset \Xbold$ of items from the stream, such that:%
\begin{equation}\label{eq:stream-b-div}
\Xobold_b^* = \argmax_{\Xobold_b \subseteq \Xbold,\, |{\Xobold_b}| =b}\,\, \mindist(\Xobold_b).
\end{equation}
The associated distance is denoted as $\mindist^* \mydef \mindist(\Xobold_b^*)$.
\end{definition}

\subsection{Mapping to a score-based sequential selection problem}\label{sec:problem_mapping}
We see the irrevocable-choice online diversification as a sequential multiple selection process, for which inspiration can be drawn from existing algorithms in that literature, such as those for the standard \SPfull \cite{Lindley61, Dynkin63}. %
The process is divided in selection \emph{rounds}, $k=1,...,b$, where each round picks one item and terminates (stopping time). Our approach is rank-based within each round, hence non-parametric, and does not make assumptions about the data distribution. The ranks are derived by associating to each incoming item a \emph{score} representing its worth %
for the selection. %

Contrary to %
standard sequential selection problems, in this setting there is no inherent fixed score that reflects the global worth of an item, since this depends on the round, thus it is essentially relative to what the selection set contains already. %
\begin{definition}\label{def:step-wise-score}
\textbf{Step-wise score function.} %
The step-wise score function $S$ evaluates the $j$-th arriving item in the stream with respect to the current selection set $\Xobold_k$ as follows:
\begin{align}\label{eq:score}
\!\!\!\!\!\!\!\!\!\!\!S_{j,k} \mydef S(x_j,\, \Xobold_k) =\ & \mindist\big(x_j,\, \Xobold_k\big)\in \real_+, \\ &\, \, 
\forall j \le N,\,\,\forall k\ \,\st\ 1 \leq k \le b.\nonumber
\end{align}
\end{definition}%
For $k \geq 2$ selected items, we can define the score of the $l$-th of those ($x_{\m_{l}}$) %
to be its minimum distance to the other selected items: %
\begin{align}\label{eq:score_selection}
\!\symb{S}_{\m_{l},k} = \ &\mindist\big(x_{\m_{l}},\, \Xobold_k \backslash x_{\m_{l}}\big)\in \real_+, \\ &\, \, \forall l \le k,\, \forall k\ \,\st\ 2 \leq k \le b.\!\!\!\!\!\!\!\nonumber%
\end{align}
At the end of the process, the selected items in $\Xobold_b$ have the scores $\Sbold = (\symb{S}_{l,b})_{l\le b}$. The quality of the selection is measured by the \emph{reward} $\phi$ that is non-increasing with $b$: %
\begin{align}
\rsymb(\Sbold%
) &=\!\! %
\min_{1\le l \le b}\symb{S}_{m_l,b}
= \mindist(\Xobold_b).
\label{eq:cost_general2}
\end{align}
The last expression goes back to \Eq{eq:stream-b-div}. Our concrete objective is thus to maximize the expectation of the reward: $\Exp{\rsymb(\Sbold%
)}$.

\begin{remark}\label{remark:first_instance} 
\emph{\textbf{Selection of the first item.}}~\textup{%
At the beginning of the process, the selection set is empty. Lacking a quality reference, \Eq{def:step-wise-score} cannot be evaluated to assign a score to an incoming item. To overcome this problem, we always select the first incoming item (\ie $\m_1=1$). Despite being a pure random selection, we empirically observed that it had little effect in the quality of the final selection, even for generally small selection sets. Note that the same choice, and on similar empirical basis, was made in \cite{Zhu16} (see Sec.\,4.1.1 therein). %
} %
\hfill\myEndBlock
\end{remark}

\begin{remark}\label{remark:iid_instance}
\emph{\textbf{I.i.d. scores within each round.}}~\textup{In our setting, the items, $x_j\in\real^d$, $j=1,...,N$, are assumed to be \iid. Then, while $\Xobold_{k}$ remains fixed during a round, the observed scores of the incoming items,
$S_{j,k} \mydef \mindist\big(x_j,\, \Xobold_k\big)$, $\forall j$ (see \Definition{def:step-wise-score}), 
are also \iid At the end of a round the selection grows, $\Xobold_{k} \subset \Xobold_{k+1}$, therefore, the scores observed in different rounds are neither independent nor identically distributed. However, having \iid scores within each round allows the analysis using the typical sequential selection framework.
}
\hfill\myEndBlock
\end{remark}

\begin{remark}\label{remark:rank_vs_scores}
\emph{\textbf{Working with ranks \vs scores.}}~\textup{The rank-based analysis we consider in each round, is essentially equivalent to a score-based approach dealing with a uniform score distribution over $[0,1]$ (see \cite{Bearden06}). This means that both approaches use the rank of the scores (the first implicitly, the second explicitly) and therefore arrive to the same selection.
}
\hfill\myEndBlock
\end{remark}
\subsection{Failures}
The constraints (C1-C3) listed in \Sec{sec:diversification_problem} %
imply that the selection process may suffer from \emph{failures}. Especially in the \MaxMin $b$-diversification problem, even one badly selected item is likely to lower drastically the selection set quality. 
\begin{definition}\textbf{Failure and failure rate.} A failure at step $j \geq N-b$, with $k<b$ items already selected, is the event of picking by default the ($b-k$)-th incoming item in order to complete the selection set, despite being below a selection threshold $\tau_j$:
\begin{equation}\label{def:failure}
f_{j} = \ind\{N-j+1 = b - k\} \,\land\,\ind\{\Xbold_j < \tau_j\}. %
\end{equation} 
The failure rate in $T$ applications of an algorithm is given by:
\begin{equation}
\frate = \frac{1}{T} \sum_{\textup{t}=1}^T \ind\big\{\textstyle\! \sum_{j=N-b}^{N} f_j(\textup{t}) > 0\big\}. 
\end{equation}%
\end{definition}

\section{Failure Rate Minimization}%
\label{sec:adjusted_algorithm}
\subsection{Principle}\label{sec:principle}
In this section, we present an overview of the \emph{\FRMfull} (\FRM) algorithm that produces high quality selection sets from a data stream, while reducing the failure rate. \Alg{alg:FRM_all} is the main \emph{multi-round} procedure, which calls \Alg{alg:FRM_round} to handle each round. Both algorithms take two types of input: a fixed one comprising the length $N$ of the stream, the number of items $b$ to select, the step-wise score function and the decremental function $q$ that controls the adaptive acceptance threshold, which we describe in the next section, and a sequential one that corresponds to the incoming items of the stream, or a subsequenceof it, respectively.

\subtitle{Multiple rounds.}~%
\Alg{alg:FRM_all} %
returns the decision vector, the selection set, and the value of the \MaxMin objective (\Eq{eq:cost_general2}). 
The first round is special as it selects automatically the first item of the stream (\ie $m_1=1$), which is purely random, yet it secures immediately a reference item for computing distance-based scores (see \Remark{remark:first_instance}).
Then, \FRM considers an initial split of the rest ($N\!-\!1$)-length stream into $b-1$ subsequences of length $n^* = \big\lfloor \frac{N-1}{b-1} \big\rfloor$ each, and performs a selection \emph{round} to pick \emph{exactly} one item from each of them. %
Each round examines at least $n^*$ items, but the precise length is determined as the procedure goes (see lines $4$, $8$, and $13$ in \Alg{alg:FRM_all}). This aspect affects the failure rate, as we discuss later in \Sec{sec:low-failures}.

\begin{algorithm}[t]
\footnotesize 
\caption{The multi-round \FRM algorithm}%
{\bf Input:}~\emph{fixed}:~the length $N$ of the stream to process; the number of items $b$ to select; the step-wise score function $S$; the decremental function $\dec;$~{--}~\emph{sequential}:~the incoming data items $\{x_1,...,x_N\}$.%

{\bf Output:} the final decision vector $\Abold{} = (A_{1},\mydots,A_{N})$ for the stream; the final selection set $\Xobold$; the value of the objective function $\phi$.%
\begin{algorithmic}[1]
\phase{Initialization and first selection (round)}\vspace{-0.7mm}%
\State {$A_1 \leftarrow 1$} \Comment{select automatically the first item of the stream}
\State $\Xobold \leftarrow \{ x_1 \}$ 
\State $\phi \leftarrow \infty$
\State $n^* \leftarrow \lfloor (N-1) / (b-1) \rfloor$ \Comment{the fixed round length}
\State $\delta \leftarrow 0$ \Comment{the number of remaining items from the prev. round}
\State $i \leftarrow 2$ \Comment{starting index of the next round\!}
\phase{Rounds 2 to b}\vspace{-0.7mm}%
\For{$k = 2$ to $b$}
\State \label{alg1:line:n} $n \leftarrow n^* + \delta$ \Comment{compute the length of the round}
\State $\{$%
$\xo{}$, $j$, $S_k$\} $\leftarrow$ {\scriptsize\texttt{OneRoundFRM}}\,$(\Xobold$, $n$, $\dec$, $\S{}$, $\{x_i,\!...\})$ \Comment{use \Alg{alg:FRM_round}}
\State $A_{i+j-1} \leftarrow 1$ \Comment{note the decision%
}
\State $\Xobold \leftarrow \Xobold{} \cup \xo{}$ \Comment{increase the selection set}
\State $\phi \leftarrow \min(\phi, S_k)$ \Comment{update the value of the objective function}
\State $\delta \leftarrow n-j$ \Comment{note the overflow of the round}\label{alg1:line:delta}
\State $i \leftarrow i+j+1$ \Comment{move to the starting position of the next round}
\EndFor%
\State \Return $\Abold{}$, $\Xobold$, $\phi$

\end{algorithmic}
\label{alg:FRM_all}
\end{algorithm}

\bigskip
\begin{algorithm}[t]
\footnotesize 
\caption{One round of the \FRM algorithm}%
{\bf Input:}~\emph{fixed}:~the length $n$ of the subsequence to process; the set $\Xobold$ of the previously selected items; the step-wise score function $S$; the decremental function $\dec$;%
~{--}~\emph{sequential}:~the incoming data items $\{x_1,...,x_n\}$.%

{\bf Output:} %
the selected item $x_j$, its index $j$ in the subsequence, and the associated score $S_j$ that is the maximized objective.%
\begin{algorithmic}[1]
\phase{Learning phase}\vspace{-0.7mm}%
\State {$c \leftarrow \lfloor \sqrt{n}-1 \rfloor$} \Comment{cutoff value: reject automatically the $c$ first items}
\State $(\S{1},...,\S{c}) \leftarrow (\S{}(x_1,\Xobold),...,\S{}(x_c,\Xobold))$ \Comment{compute the scores, \Eq{eq:score}}
\State $\Xrefbold{c} \leftarrow \text{sort}(\S{1},\mydots,\S{c})$ \Comment{order from the best to the worst score}
\State $\thresh \leftarrow \Xref{1}{c}$ \Comment{set the fixed threshold to the best score}
\phase{Selection phase}\vspace{-0.7mm}%
\For{$j = c+1$ to $n$}
\State $\S{j} \leftarrow \S{}(x_j,\Xobold)$ \Comment{compute the item's score, \Eq{eq:score}}
\If { ${\mu}_j - {\sigma}_j > 0$ } \Comment{see \Proposition{prop:mu}}
\State {$\thresh_j \leftarrow \Xrefbold{q_j,j}$} \Comment{use the adaptive threshold, \Eq{eq:q_j_exp}}
\Else
\State $\thresh_j \leftarrow \thresh$ \Comment{use the fixed threshold}%
\EndIf
\If {$\S{j} > \thresh_j$ or\,\,\,$ j == n$} \Comment{if $x_j$ beats the threshold, or failure}
\State {\textbf{break}} \Comment{terminate the round}
\Else 
\State $\Xrefbold{j} \leftarrow {\scriptsize\texttt{sort}}(\Xrefbold{j-1} \cup \S{j})$ \Comment{update the ordered set}
\EndIf
\EndFor%
\State \Return 
$x_j$, $j$, $S_j$%
\Comment{$j$ is the stopping position}
\end{algorithmic}
\label{alg:FRM_round}
\end{algorithm}

\vspace{-1em}
\subtitle{Single round.} This concerns how \FRM operates in each round $k \in \{2,\mydots,b\}$, hence we omit the index $k$ from the notations. 
\Alg{alg:FRM_round} operates on a subsequence (substream) of length $n$ by taking into account the items selected by the previous rounds. Each item $x_i$, $\forall i =1,\mydots,n$, is examined one by one, and the round terminates with the selection of -say- the $j$-th item. As the last $n-j$ items of the subsequence are not examined, we call them \emph{overflow} of the round. %
\FRM uses a logic similar to the standard algorithm for the \SPfull \cite{Dynkin63}. More specifically, each subsequence is further divided in two phases: the \emph{learning phase} of length equal to a \emph{cutoff} value $c \in \mathbbm{N}^*$ (see lines $1$-$4$ in \Alg{alg:FRM_round}), and the \emph{selection phase} (lines $5$-$18$ in \Alg{alg:FRM_round}). %
In the former, all $c$ items are rejected and the best (\eg in the second round, that will be the farthest away distance to the first selected item) is recorded as: 
$\tau = \max_{j=1,...,c}\{S_j\} \in \real_+$.
The value $\tau$ is then used as an \emph{acceptance threshold} for actually picking an item in the selection phase (see line $3$ in \Alg{alg:FRM_round}). 
The round terminates when an item is selected, \ie when for some index $j\in \{c+1,\mydots,n\}$, $S_{j} > \thresh$, or at the end of the subsequence if none managed to beat the acceptance threshold. In the latter case, the last item of the subsequence is selected by default, signifying a failure.

The optimal size for the learning phase depends on the length of the subsequence and the objective function under consideration. %
For items with numerical scores that are \iid uniformly distributed in $[0,1]$, and the objective to be the maximization of the score of the selected item, the optimal size has been shown to be $c = \sqrt{n}-1$ \cite{Bearden06}. This tuning holds also for the rank-based settings due to its equivalence to the latter score-based setting (see \Remark{remark:rank_vs_scores}).

\subsection{The low-failures adjustment}\label{sec:low-failures}
\FRM mitigates the effects of failures by: i)~defining a \emph{switch position} $\switch$ in the subsequence, at which the acceptance threshold changes from static to dynamic, and ii)~by carefully redistributing the overflow to the rest of the rounds. %
\begin{definition}\textbf{Switch position.} %
The switch position is the index $\switch \in \{c+1,\mydots,n\}$ in the subsequence at which the acceptance threshold $\thresh_j$ turns from a static ($\tau$) to a dynamic one ($\thresh^\textup{upd}_j$) that gets updated from the $j^*$-th step and on: 
\begin{equation}
{\thresh}_j \mydef 
\begin{cases}
\thresh & j < \switch; \\
\thresh^\textup{upd}_j & \textup{otherwise}.%
\end{cases}
\label{eq:adj_threshold}
\end{equation}
\end{definition}
There are three questions arising naturally. %
Next, we answer the first question (Q.1) analytically (the technical proofs are provided in the Appendix), while the answers to the other two (Q.2 \& Q.3) result from our empirical study.

\medskip
\noindent\subtitle{Q.1. How to choose an appropriate switch position $j^*$?}\\
We compute the expected stopping time assuming that no failure has occurred, and that the horizon $n$ is known and finite. %
Since selecting an item means ignoring the rest of the sequence and putting an end to the round, this is equivalent to investigating the expectation of the decision variable $\tilde{A}_j \mydef \aj \in \{0,1\}$, where $A_j \in \{0,1\}$, $\forall j$. 
Formally, we denote this by ${\mu}_j(c) \mydef \Exp{\tilde{A}_j \mid \textup{no fail.}}$, $\forall j$, which is computed as \Proposition{prop:mu} describes.
\begin{proposition} Let a pool of $n$ %
totally ordered items, %
ranked from the best ($1$-st) to the worst ($n$-th). The expected %
global rank of the best item %
occurring in a %
random subset of size $c$ items from the pool, such that $1 \leq c \leq n$, is given by: 
\begin{equation}\label{eq:gamma_1}
\gamma(c) = \frac{n+1}{c+1}.
\end{equation}
\label{prop:gamma}
\end{proposition}
\begin{proposition} Provided that no failure occurs at the end of the subsequence, the expectation of having selected an item %
at step $j > c$ is given by:
\begin{align}
\label{eq:mu}
{\mu}_j(c) &= \displaystyle \frac{1-\frac{\binom{n-(\gamma-1)}{j-c}}{\binom{n}{j-c}}}{1-\frac{\binom{n-(\gamma-1)}{n-c}}{\binom{n}{n-c}}} \ \ \in [0,1],%
\end{align}
where 
$\gamma \mydef \gamma(c)$ is given by \Eq{eq:gamma_1}. Then, the standard deviation is by its definition: $\sigma_j(c) = \sqrt{ {\mu}_j(c) - {{\mu}_j^2(c)}}$.
\label{prop:mu}
\end{proposition}

We can use \Proposition{prop:mu} with $c=\sqrt{n}-1$, to compute ${\mu}_{j} \mydef {\mu}_{j}(c)$ and $\sigma_j \mydef {\sigma}_{j}(c)$, for any step $j \in \INTerval{c+1}{n}$.
Then, a simple mechanism for preventing failures is to compare the current decision variable with its expectation, which is expressed by the following rule: %
\emph{if $\tilde{A}_j$ is less than $\sigma_j$-lower from its expected value $\mu_j$}, \emph{the threshold should be relaxed}. Formally: ${\mu}_j-{\sigma}_j > \tilde{A}_j\xRightarrow[]{} {\mu}_j-{\sigma}_j > 0$%
, since $\tilde{A}_j$ in our setting is always equal to zero until an item is picked. The next proposition states that the switch position of \Definition{eq:adj_threshold} exists and is unique. %
\begin{proposition} In our setting, the switch position $j^*$ of \Definition{eq:adj_threshold} exists and is unique, such that:%
\begin{equation}
\\ \phantom{xxxxx} \exists! \, j^* \in [c+1,n]\,:\ \ %
\begin{cases}
\mu_{j}-\sigma_{j} \leq 0, \ \ j < j^*;\\
\mu_{j}-\sigma_{j} > 0, \ \ j \geq j^*.
\end{cases}
\end{equation}
\label{prop:existence_j*}
\end{proposition}

\noindent\subtitle{Q.2. How to dynamically adjust the threshold for $j>j^*$?} \\
Let the descending list of scores up to the $j$-th item of the subsequence be denoted by $\Xrefbold{j} = (\Xref{1}{j}, \mydots,\Xref{j}{j})$. Our strategy adjusts the threshold by introducing the decremental term $\dec_j \mydef q(j) : \INTerval{c+1}{n} \rightarrow \mathbbm{N}^*$, which is a function of the position $j$. %
Before the switch $j^*$, the fixed threshold $\thresh$ is the score at the index $1$ in the list $\Xrefbold{j}$, with no decrementation, \ie $\dec_j=0$.
After the switch $j^*$, the threshold $\thresh^\textup{upd}_j$ is the value in $\Xrefbold{j}$ whose index is $\dec_j$ (lines $4$, $9$, and $12$ in \Alg{alg:FRM_round}). %
More concretely, the step-wise threshold explicits \Eq{eq:adj_threshold} as follows:%
\begin{equation}
{\thresh}_j \mydef 
\begin{cases}
\displaystyle \tau = \Xref{0}{j} & j < \switch \\
\thresh^\textup{upd}_j = \Xref{\dec_j}{j} & \textup{otherwise}. \\
\end{cases}
\label{eq:adj_threshold_final}
\end{equation}

There is a lot of room for designing decremental functions. In this work, we use a heuristic computed by the following expression, involving the current item index $j$ and two hyperparameters:
\begin{equation}\label{eq:q_j_exp}
\dec_j = \big\lfloor\alpha \cdot j \cdot \big({\textstyle\frac{j}{n}})^\beta\big\rfloor, \ \ \alpha \in[0,1], \ \beta \geq 1.
\end {equation}
The hyperparameters $\alpha$ and $\beta$ control respectively the maximum allowed decrement and the decay rate of the decrement. This form imposes a mild relaxation at early steps, which becomes stronger when approaching the end of the sequence. %
\medskip
\noindent\subtitle{Q.3. How to manage the overflow of each round?}\\
When a round terminates by selecting an item at position $j$, there is an overflow of $\delta = n-j$ items that are not going to be examined, thus are not at all consumed from the stream. Instead of wasting that part, it is natural to seek ways to best use it in later rounds. $\delta$ is independent to the round index $k$, and is expected to be larger as $n$ gets larger%
. Moreover, the longer a round is, the better its expected selection quality.

Overflow management essentially decides which of the next rounds are worth to become longer by assigning to them a part of the overflow. In practice, this is not about asigning the data items per se, but mostly about how many items will be consumed from the stream by each next round. This is a subject that requires deeper theoretical analysis for revealing the relation of the main variables of interest. Nevertheless, in this work we follow a simple approach that passes the overflow of a round $k$ to the right next to it, for any $k>1$ (lines $8$ and $13$ in \Alg{alg:FRM_all}). 
Receiving overflow makes round $k+1$ longer, and its overflow also bigger in expectetion. Therefore, passing over the overflow implies later rounds will accumulate more length, as long as the selection process is smooth and there are no failures. However, when a failure occurs in a round, it implies there will be no overflow to pass over to its subsequent rounds ($\delta=0$).
\section{Relevant algorithms from the literature}\label{sec:diff_algo}

\KLEINBERG is a well-known $k$-choice algorithm, which operates recursively and does not employ a fixed-length learning phase, %
as is done in the classical secretary algorithm. Other algorithms, namely \OPTIMISTIC \cite{Babaioff07}, \VIRTUAL \cite{Babaioff07}, \SREF \cite{Albers19}, and the more recent \VIRTUALp \cite{Mla22}, follow a two-phase structure: a learning phase, where the first $c=N/e$ items are discarded to compute selection thresholds, and then a selection phase. The main distinction among these lies in how candidates are evaluated and selected during the selection phase, based on different thresholding strategies. 

We also compare with \MEAN and \MEDIAN \cite{Broder09}, which are designed for a hiring problem without two-phase structure, whose acceptance threshold is the average/median score of the items selected so far up to the current step in the sequence. In all of these variants, an item does not have a fixed score as mentioned; instead, the score is computed by a step-wise score function within the algorithm. Additionally, when the end of the stream is reached and the selected set is not yet complete, we automatically select the remaining items, even though this mechanism is not part of the original algorithm design. 

Finally, we have \SIMPLEK and \DynSIMPLEK \cite{Zhu16}, both these last two use almost the same amount of resources as an offline procedure (\ie they store and process offline half of the stream, \ie $c=\frac{1}{2}N$, and do a binary search during the learning phase). Their main difference is the dynamic thresholding of \DynSIMPLEK that aims at reducing the failure rates and thus is the most natural competitor in terms of objective. As this is a point in common with our \FRM, we include it in the comparative results as a sort of \emph{oracle} baseline. %
\section{Experiments}\label{sec:exps}
\subsection{Setup}

\Tab{tab:failures_mindist} lists all the compared selection strategies and their features (see also \Sec{sec:diff_algo}). The strategies are of three types: those that are online, \SIMPLEK and \DynSIMPLEK that are hybrid/offline since they store and process a large part of the stream, and \GREEDY that is an offline. Among those, only the two \SIMPLEK variants and the proposed \FRM (and its variants) %
address specifically the streaming diversification problem, %
while also minimizing the rate of selection failures $\frate$. 
As a way to assess the contribution of the different components of the \FRM algorithm, we additionally test the \FRMbasic variant, where the threshold decrement and overflow management are not implemented, and the \FRMnooverflow variant which adjusts the threshold dynamically but has no overflow management. Moreover, we use the decremental function $q_j$ of \Eq{eq:q_j_exp} with $\alpha=0.1, \beta=2$, but note that simpler. Finally, we use as reference the \GREEDY algorithm that is purely offline: it selects the first item in the sequence (to comply with what \FRM does) and then uses the whole batch of the items and the \MaxMin objective to build incrementally the selection set.%

Regarding the datasets we use in the evaluation, we firstly consider the scenario \Synth$_1$ by generating uniformly synthetic streams. Each stream comprises $N = 5000$ items in $d = 2$ dimensions, specifically bounded in $[0,1]^d$. 
\Synth$_2$ is also a synthetic scenario, having streams with items generated by a uniform distribution, this time in $d = 3$ dimensions. 
We choose to use low-dimensional items to prevent the curse of dimensionality from interfering with the distance calculations, as distances become less informative in high-dimensional spaces. The use of long streams reflects real-world conditions, where streams are typically long, like the following rocks data. More specifically, the real dataset provided in \cite{Eco1} \cite{Aster2} concerns a robotic sampling scenario. Out of the $1189682$ rocks, represented in $d = 2$ dimensions, that this dataset contains in total, we draw two random samples to create two scenarios: Rocks$_1$ and Rocks$_2$, which have respectively $N = 10000$ and $N = 5000$ rocks. Random streams are generated out the associated sample in each case.%

\begin{figure*}[t!]
\centering
\begin{subfigure}[b]{0.49\textwidth}
\centering
\clipbox{0pt 0.45pt 0pt 0pt}{\includegraphics[width = 1\linewidth]{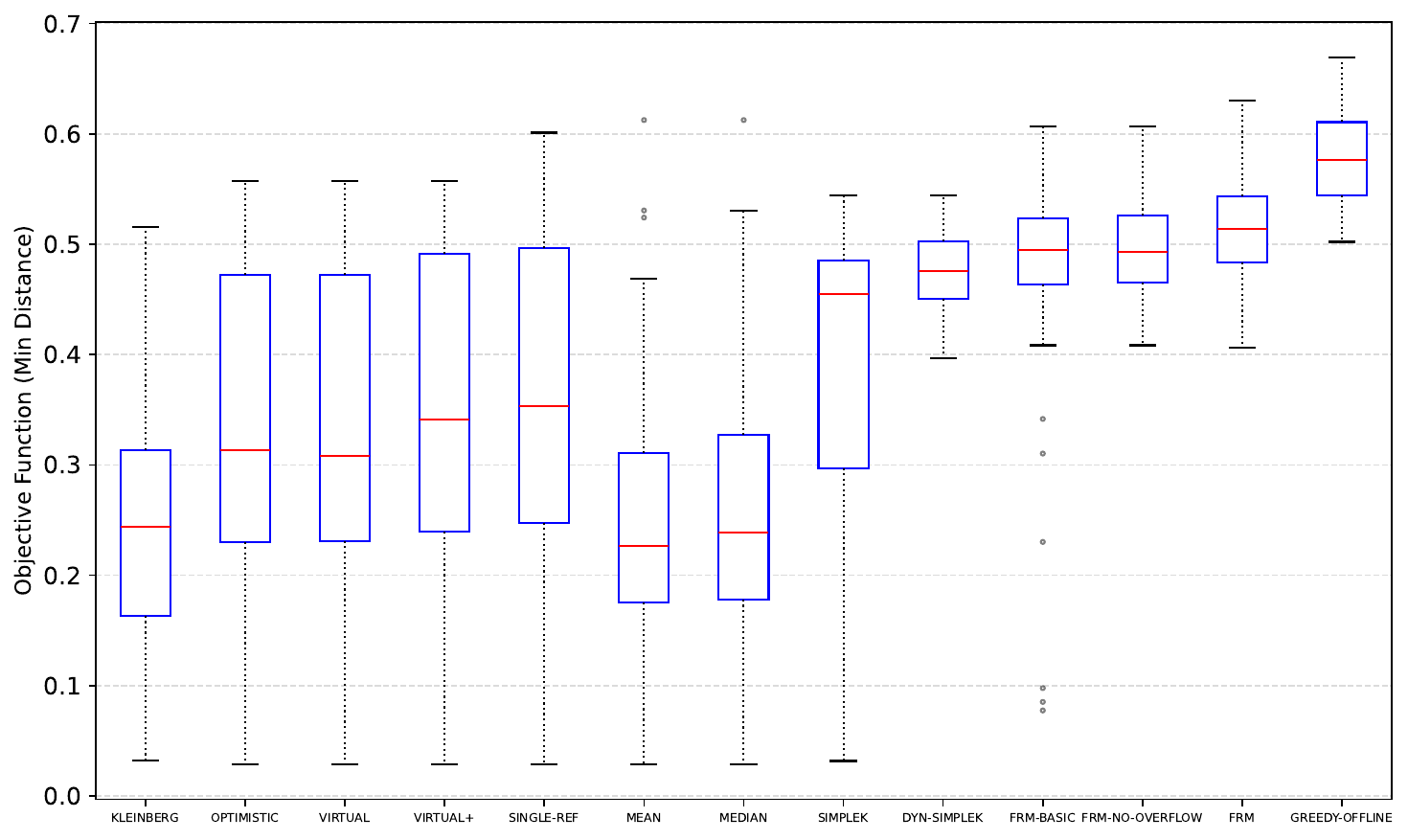}}
\caption{\Synth$_1$}
\label{fig:syn}
\end{subfigure}
~
\begin{subfigure}[b]{0.49\textwidth}
\centering
\clipbox{0pt 0.45pt 0pt 0pt}{\includegraphics[width = 0.99\linewidth]{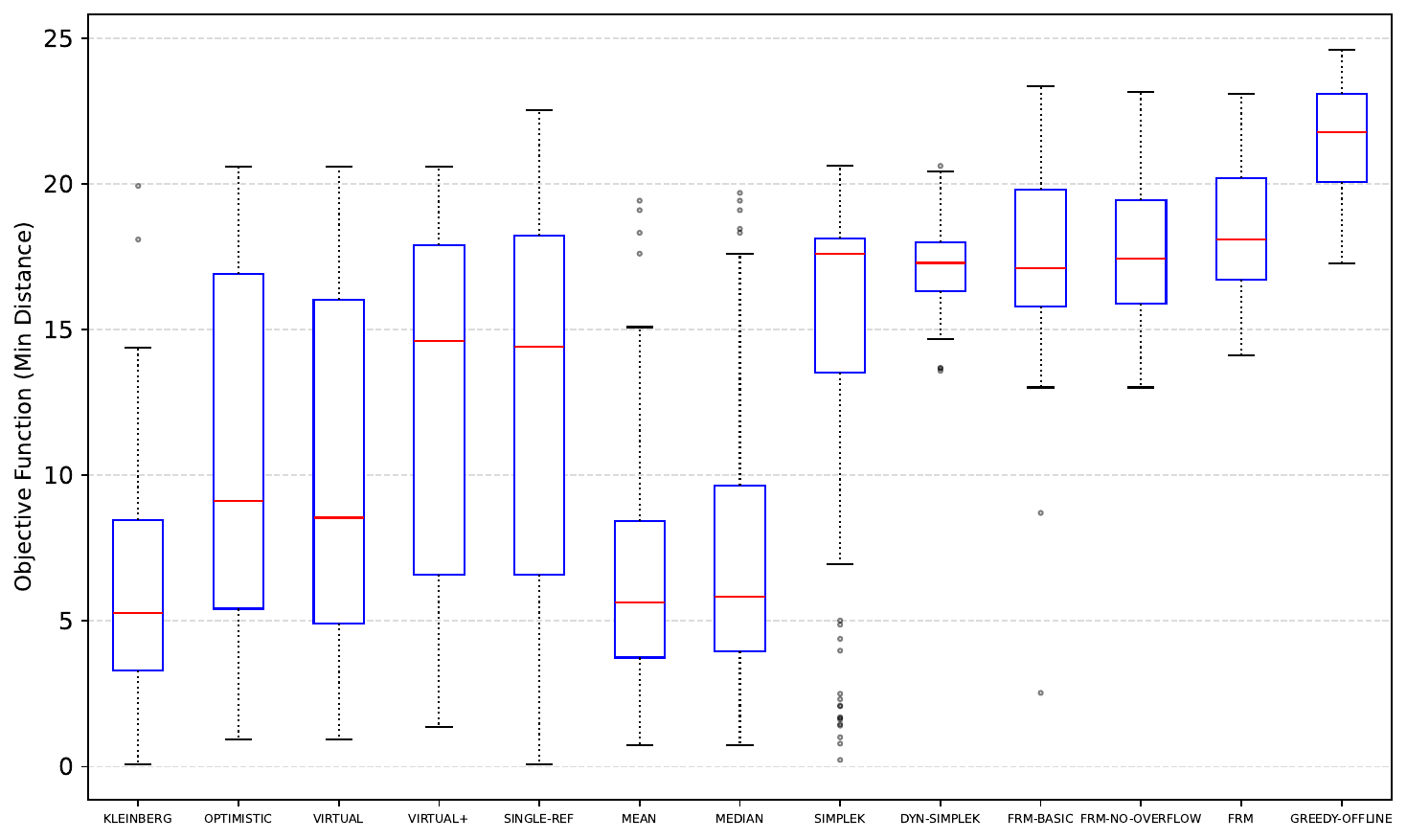}}
\caption{Rocks$_1$}
\label{fig:real}
\end{subfigure}
\vspace{-5mm}
\caption{Minimum distance between the $b=5$ selected items from (a) synthetic streams of the \Synth$_1$ scenario, and (b) real rocks. In both cases the streams contain $N=5000$ items in $d=2$ dimensions. The results are averaged over $100$ tests.
}
\label{fig:barplots}
\vspace{1em}
\end{figure*}

\begin{figure*}[t!]
\centering
\begin{subfigure}[b]{0.49\textwidth}
\centering
\clipbox{0pt 0.45pt 0pt 0pt}{\includegraphics[width = 1\linewidth]{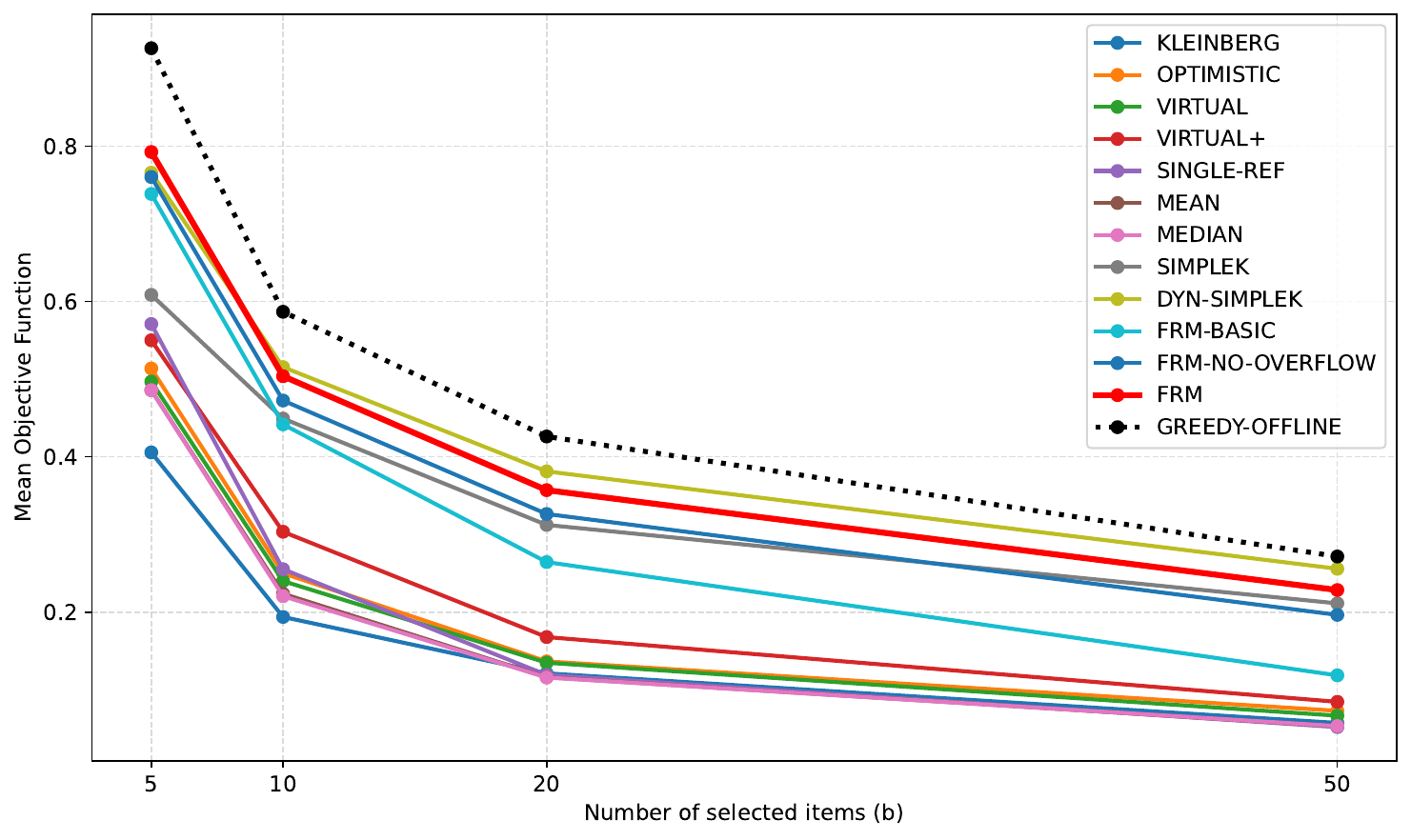}}
\caption{\Synth$_2$}
\label{fig:comp_syn}
\end{subfigure}
~
\begin{subfigure}[b]{0.49\textwidth}
\centering
\clipbox{0pt 0.45pt 0pt 0pt}{\includegraphics[width = 0.99\linewidth]{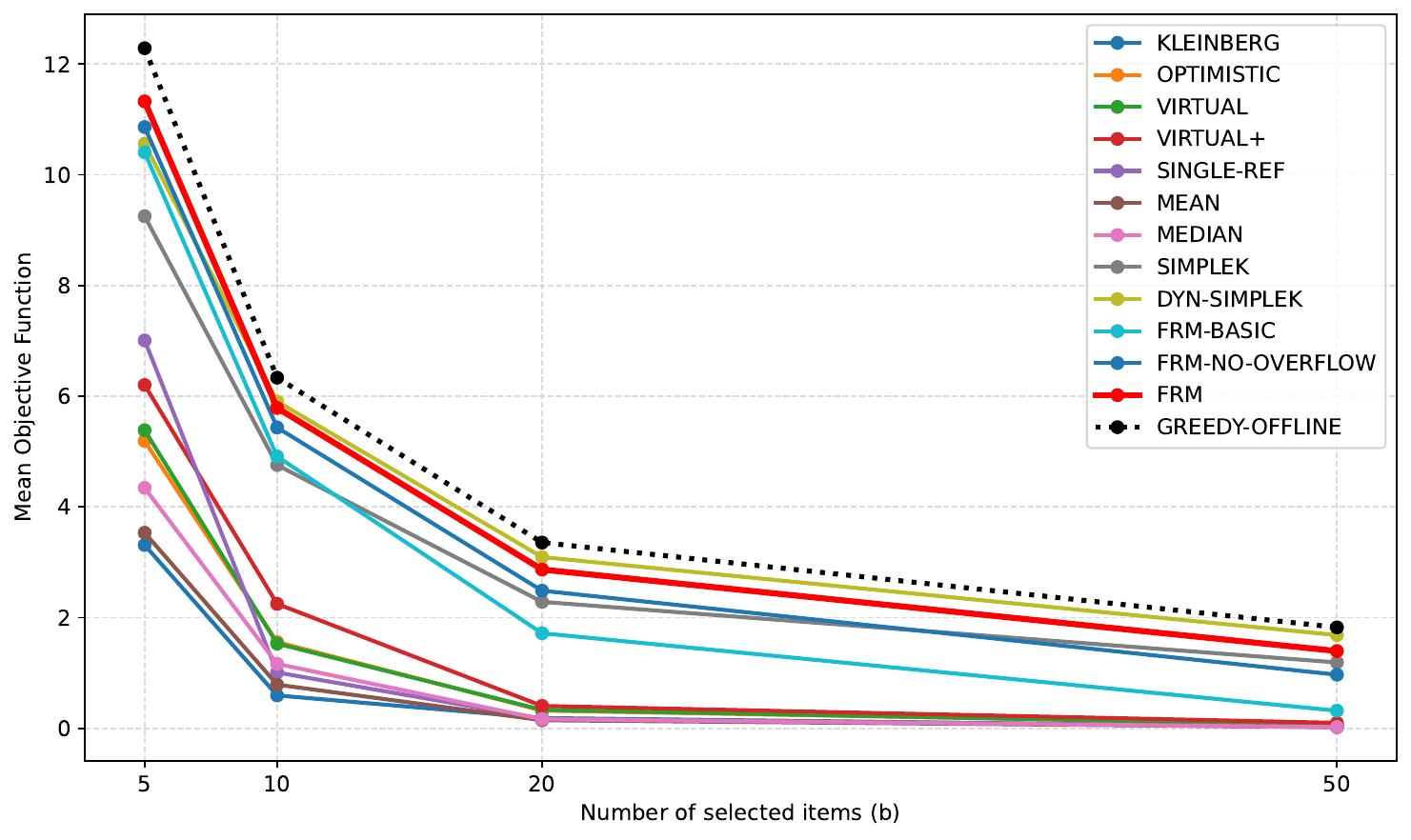}}
\caption{Rocks$_2$}
\label{fig:comp_real}
\end{subfigure}
\vspace{-5mm}
\caption{Minimum distance between the $b=5, 10, 20, 50$ selected items from (a) synthetic streams of the \Synth$_2$ scenario ($N=5000$ items in $d=3$ dimensions), and (b) real rocks ($N=10000$ items in $d=2$ dimensions). The results are averaged over $100$ tests. Continuous lines correspond to online selection strategies; dashed lines are largely offline strategies that learn on a big part of the stream before starting the selection; the dotted black line is an offline approximation that we use as reference.
}
\label{fig:lineplots}
\vspace{1em}
\end{figure*}

\renewcommand{\toprule}{\noalign{\hrule height 1.2pt}}
\renewcommand{\bottomrule}{\noalign{\hrule height 1.2pt}}
\renewcommand{\midrule}{\noalign{\hrule height 0.9pt}}

\begin{table*}[h!]
\centering
\resizebox{0.95\textwidth}{!}{
\begin{tabular}{l c c c | r r r r}
\toprule
&&&& \multicolumn{4}{c}{\textbf{Datasets}}\\
\textbf{Selection strategy} & \textbf{Online} & \textbf{Diversifies} & \textbf{Minimizes} $\frate$ & \textbf{\Synth}$_1$ ($b=5$) & \textbf{\Synth}$_2$ ($b=5$)& \textbf{Rocks}$_1$ ($b=5$)&\textbf{Rocks}$_2$ ($b=5$) \\
&&&& $\frate\sep\mindist$\phantom{xx} & $\frate\sep\mindist$\phantom{xx} & $\frate\sep\mindist$\phantom{xx} & $\frate\sep\mindist$\phantom{xx} \\
\midrule
\KLEINBERG & \checkmark &&& 2.97\sep0.25 & 2.78\sep 0.41 & 3.11\sep\phantom{0}6.12 & 3.03\sep3.31\\ 
\OPTIMISTIC & \checkmark &&& 0.68\sep0.33 & 0.83\sep 0.51 & 0.70\sep10.48 & 0.98\sep5.20 \\ 
\VIRTUAL & \checkmark &&& 0.70\sep0.33 & 0.89\sep 0.50 & 0.73\sep10.10 & 0.98\sep5.38 \\ 
\VIRTUALp & \checkmark &&& 0.60\sep0.35 & 0.72\sep 0.55 & 0.52\sep12.56 & 0.82\sep6.20 \\ 
\SREF & \checkmark &&& 0.60\sep0.36 & 0.72\sep 0.57 & 0.57\sep12.52 & 0.74\sep7.01 \\ 
\MEAN & \checkmark &&& 1.12\sep0.25 & 0.45\sep 0.49 & 1.06\sep\phantom{0}6.45 & 1.24\sep3.53 \\ 
\MEDIAN & \checkmark &&& 0.97\sep0.26 & 0.44\sep 0.49 & 0.80\sep\phantom{0}7.26 & 0.85\sep4.34 \\ 
\SIMPLEK ($c=.5N$) & & \checkmark & \checkmark & 0.01\sep0.39 & 0.04\sep 0.61 & 0.00\sep14.66 & 0.01\sep9.25 \\ 
\DynSIMPLEK ($c=0.5N$) & & \checkmark & \checkmark & 0.00\sep0.47 & 0.00\sep 0.77 & 0.00\sep17.22 & 0.00 10.56 \\ 
\hdashline
\FRMbasic & \checkmark & \checkmark & \checkmark & 0.09\sep0.48 & 0.14\sep 0.74 & 0.07\sep17.57 & 0.10 10.41 \\ 
\FRMnooverflow & \checkmark & \checkmark & \checkmark & 0.00\sep0.50 & 0.0\sep 0.76 & 0.00\sep17.73 & 0.00 10.86 \\ 
\FRM & \checkmark & \checkmark & \checkmark & \textbf{0.00\sep0.52} & \textbf{0.0\sep 0.79} & \textbf{0.00\sep18.36} & \textbf{0.00 11.33} \\ 
\midrule
\GREEDY & & \checkmark & \checkmark & N/A\sep0.58 & N/A\sep 0.93 & N/A\sep21.56 & N/A 12.29 \\ 

\bottomrule
\end{tabular}
}
\vspace{1mm}
\caption{The failure rate $\frate$ (lower is better) in \% and the median minimum distance $\mindist$ (higher is better) achieved by the compared algorithms (rows) in streaming \MaxMin $b$-diversification tasks on two synthetic and two real data streams. In all cases, the selection size is fixed at $b=5$. The best results, among the algorithms that are online or have an online selection phase, are highlighted. The last row corresponds to an offline reference algorithm.}
\label{tab:failures_mindist}
\end{table*}

\subsection{Results}

\Fig{fig:barplots} displays the minimum distance, averaged over $100$ tests, among the $b = 5$ items of a selection made by each of the compared strategies. The two subfigures present the results for \Synth$_1$ and Rocks$_1$ cases.
For each strategy, the red line represents the median performance, the blue box is the $25$-th and $75$-th percentiles, the whiskers extend to the most extreme data points that are not outliers, and the outliers are shown with black dots. 
In \Fig{fig:syn}, \FRM demonstrates the best overall performance in terms of both mean and median, with its upper quartile even approaching the performance of the \GREEDY offline baseline. Additionally, the performance gap between \FRM and \FRMnooverflow highlights the benefits of the overflow mechanism. Finally, the presence of outliers in \FRMbasic, compared to \FRMnooverflow, suggests that the adaptive threshold contributes to better control over the quality of the selection process. 
The results in \Fig{fig:real}, which refer to real data for which no prior knowledge of the data distribution (or range of values) is known, we observe a similar trend as previously: \FRM has consistently the best overall performance, even for longer streams.

\Fig{fig:lineplots} displays the minimum distance among the $b = 5, 10, 20, 50$ selected items, achieved by the algorithms, averaged over $100$ tests. The two subfigures refer to streams of \Synth$_2$ and Rocks$_2$ scenarios.
In \Fig{fig:comp_syn}, we can remark that the performance of the algorithms tends to converge when $b$ increases, while \FRM(red line) and \DynSIMPLEK(blue dash) are having always the best performance. The results on real data shown in \Fig{fig:comp_real} are similar to those on synthetic data.

Finally, \Tab{tab:failures_mindist} reports the failure rate and the value of the \MaxMin diversification objective when $b=5$. In all cases, \FRM achieves the highest mean of minimum distance, reaching approximately in each case $\frac{0.52}{0.58} \approx 90\%$, $\frac{0.79}{0.93} \approx 85\%$, $\frac{18.36}{21.56}\approx 85\%$, and $\frac{11.33}{12.29} \approx 92\%$ of \GREEDY's reference performance. Regarding the failure rate, \FRMbasic's (without adaptive threshold) failure rates are respectively $0.09$, $0.14$, $0.07$, and $0.1$, while \FRMnooverflow (with adaptive threshold) achieves a failure zero rate. This clearly demonstrates the effectiveness of the adaptive threshold in eliminating failures.

\section{Conclusion}

This paper investigated the \streambased \MaxMin $b$-diversification problem, in which the goal is to select the most diverse set possible out of the items of a finite data stream by making immediate and irrevocable decisions. %
For this problem we presented the \FRM algorithm that minimizes the selection failures that may naturally occur in this setting. The algorithm is easy-to-implement, efficient and computationally cheaper than \DynSIMPLEK that is among the main competing approaches with the main objective. 
\FRM was found to have better performance compared to relevant online algorithms of the literature through simulations on synthetic data streams and data from a real-world application of robotic sampling. Future work could investigate theoretically how to tune better the different features of the \FRM algorithm.%

\section*{Appendix}

\noindent\subtitle{Proof of \Proposition{prop:gamma}.}~
As scores are \iid within a round%
, we can obtain the probability of the global rank $R$ of the instance having a local rank $i$ in a subset of size $c$, with $r \in [1,n]$: 
$$
\Prob(R = r \mid i) = \frac{\binom{r-1}{i-1}\binom{n-r}{c-i}}{\binom{n}{c}}.
$$%
Thus, the expected global rank knowing the local rank is:
$$
\Exp{R \mid i} = \sum_{r=i}^{n-c+i} r \cdot \frac{\binom{r-1}{i-1}\binom{n-r}{c-i}}{\binom{n}{c}},
$$%
which can be simplified as follows:
$$
\small
\begin{aligned}
\Exp{R \mid i}
&=\frac{\sum_{r=i}^{n-c+i}(r-1+1)\binom{r-1}{i-1}\binom{n-r}{c-i} }{\binom{n}{c}} \\
&=\frac{\sum_{r=i}^{n-c+i}(r-1)\binom{r-1}{i-1}\binom{n-r}{c-i}+\sum_{r=i}^{n-c+i}\binom{r-1}{i-1}\binom{n-r}{c-i}}{\binom{n}{c}} \\
&=\frac{\sum_{k=i-1}^{n-c+i-1}\binom{k}{i-1}\binom{n-k-1}{c-i} + \sum_{k=i-1}^{n-c+i-1} k \binom{k}{i-1}\binom{n-k-1}{c-i}}{\binom{n}{c}} \\
&=\frac{\binom{n}{c}+ \sum_{k=i-1}^{n-c+i-1} (i\binom{k}{i}+(i-1)\binom{k}{i-1})\binom{n-k-1}{c-i} }{\binom{n}{c}} \\
&= \frac{\binom{n}{c} +i\binom{n}{c+1}+(i-1)\binom{n}{c}}{\binom{n}{c}}\\
&= \frac{i(n+1)}{c+1}.
\end{aligned}
$$
In our case, the threshold is the best item in the learning phase, hence it is sufficient to set $i=1$ to the above. \hfill\myEndBlockProof

\bigskip
\noindent\subtitle{Proof of \Proposition{prop:mu}.}~%
Given $c,n,\gamma$, let us denote $r=\gamma-1$, the number of ranks higher than the expected rank of the threshold. Thus we have $n-r$ ranks that can lead to a failure. 
The `no failure' event happens with probability : $1-\Prob(\tilde{A}_n = 0)=1-\Prob(\text{fail in } n-c)$. Analogically, $\Prob(\tilde{A}_j = 1,\, \tilde{A}_n=1) = 1-\Prob(\text{fail in} j-c)$

Thus, provided that there has occurred no failure, the expected number of selected items is: 
{\small
\begin{align}
\Exp{\tilde{A}_j \,|\, \textup{no fail.}} 
&= \frac{\Prob(\tilde{A}_j = 1,\, \tilde{A}_n=1)}{\Prob(\text{no fail.})}\nonumber%
= \frac{1-\frac{\binom{n-r}{j-c}}{\binom{n}{j-c}}}{1-\frac{\binom{n-r}{n-c}}{\binom{n}{n-c}}} \ \ \leq 1.\nonumber%
\end{align}
}%
The expression of $\sigma_j(c)$ is obtained using the fact that the decisions take specific values, $\tilde{A}_j \in \{0,1\}$, and hence $\Exp{(\tilde{A}_j)^2 \,|\, \textup{no fail.}} = \Exp{\tilde{A}_j \,|\, \textup{no fail.}} \ \leq 1$.\hfill\myEndBlockProof
\bigskip

\noindent\subtitle{Proof of \Proposition{prop:existence_j*}.}~
First, we show that $\mu_j \in [0,1]$ is increasing in terms of $j$. Recall from \Proposition{prop:mu} that: 
$$\mu_j=\frac{1-\frac{\binom{n-(\gamma-1)}{j-c}}{\binom{n}{j-c}}}{1-\frac{\binom{n-(\gamma-1)}{n-c}}{\binom{n}{n-c}}} \in [0,1].$$
Since the denominator does not depend on j, this is equivalent to showing that the nominator is increasing, \ie : $g(j) = \frac{\binom{n-(\gamma-1)}{j-c}}{\binom{n}{j-c}}$ is decreasing. 
We can easily see that, for $\gamma \geq 1$:
 $$\frac{g(j+1)}{g(j)} = \frac{[n-(\gamma-1)]-[j+1-c]}{n-[j+1-c]} \leq 1.$$ 
Besides, we have ${\mu}_j-{\sigma}_j > 0 \implies \mu_j-\sqrt{\mu_j-\mu_j^2} > 0 \implies \mu_j > {\frac{1}{2}}$, thus : 
$$j^* = \argmin_j \{\mu_j > {\textstyle\frac{1}{2}}\}.$$ 
Finally, once the item at the switch position $j^*$ is reached, the following items for any $j > j^*$ will always fall under the adaptive threshold case of \Definition{eq:adj_threshold}. Indeed, for $\mu_j \in [\frac{1}{2},1]$, ${\mu}_j-{\sigma}_j = \mu_j-\sqrt{\mu_j-\mu_j^2}$ is increasing. Thus, for all $j >j^*$, we have $\mu_j \geq \mu_{j^*}>\frac{1}{2} \implies {\mu}_j-{\sigma}_j \geq {\mu}_j^*-{\sigma}_j^* > 0 $.
\hfill\myEndBlockProof

\bigskip
\balance
\bibliographystyle{IEEEtrans} 

\begin{thebibliography}{10}
\providecommand{\url}[1]{#1}
\csname url@samestyle\endcsname
\providecommand{\newblock}{\relax}
\providecommand{\bibinfo}[2]{#2}
\providecommand{\BIBentrySTDinterwordspacing}{\spaceskip=0pt\relax}
\providecommand{\BIBentryALTinterwordstretchfactor}{4}
\providecommand{\BIBentryALTinterwordspacing}{\spaceskip=\fontdimen2\font plus
\BIBentryALTinterwordstretchfactor\fontdimen3\font minus
  \fontdimen4\font\relax}
\providecommand{\BIBforeignlanguage}[2]{{%
\expandafter\ifx\csname l@#1\endcsname\relax
\typeout{** WARNING: IEEEtran.bst: No hyphenation pattern has been}%
\typeout{** loaded for the language `#1'. Using the pattern for}%
\typeout{** the default language instead.}%
\else
\language=\csname l@#1\endcsname
\fi
#2}}
\providecommand{\BIBdecl}{\relax}
\BIBdecl

\bibitem{Vitter85}
J.~S. Vitter, ``Random sampling with a reservoir,'' \emph{ACM \Transactions on
  Mathematical Software}, vol.~11, no.~1, pp. 37--57, 1985.

\bibitem{Muthukrishnan05}
S.~Muthukrishnan, \emph{Data streams: Algorithms and applications}.\hskip 1em
  plus 0.5em minus 0.4em\relax Now Publishers Inc, 2005.

\bibitem{OODS2009}
Y.~Girdhar and G.~Dudek, ``Optimal online data sampling or how to hire the best
  secretaries,'' in \emph{\Proceedings Canadian \Conference on Computer and
  Robot Vision}, 2009, pp. 292--298.

\bibitem{OnlineNavigation2010}
------, ``Online navigation summaries,'' in \emph{\Proceedings IEEE
  \International \Conference on Robotics and Automation}, 2010, pp. 5035--5040.

\bibitem{IncrDiv2011}
E.~Minack, W.~Siberski, and W.~Nejdl, ``Incremental diversification for very
  large sets: A streaming-based approach,'' in \emph{\Proceedings ACM SIGIR
  \International \Conference on Research and Development in Information
  Retrieval}, 2011, pp. 585--594.

\bibitem{Zhu16}
Y.~Zhu and E.~J. Keogh, ``Irrevocable-choice algorithms for sampling from a
  stream,'' \emph{Data Mining and Knowledge Discovery}, vol.~30, pp. 998--1023,
  2016.

\bibitem{irrevocable-sampling-periodic-data-2018}
G.~Flaspohler, N.~Roy, and Y.~Girdhar, ``Near-optimal irrevocable sample
  selection for periodic data streams with applications to marine robotics,''
  in \emph{\Proceedings IEEE International Conference on Robotics and
  Automation}, 2018, pp. 5691--5698.

\bibitem{Chandrasekaran81}
R.~Chandrasekaran and A.~Daughety, ``Location on tree networks: $p$-centre and
  $n$-dispersion problems,'' \emph{Mathematics of Operations Research}, vol.~6,
  no.~1, pp. 50--57, 1981.

\bibitem{Akagi18}
T.~Akagi, T.~Araki, T.~Horiyama, S.-i. Nakano, Y.~Okamoto, Y.~Otachi,
  T.~Saitoh, R.~Uehara, T.~Uno, and K.~Wasa, ``Exact algorithms for the
  {M}ax-{M}in dispersion problem,'' in \emph{Frontiers in Algorithmics},
  J.~Chen and P.~Lu, Eds., 2018.

\bibitem{Ziegler05}
C.-N. Ziegler, S.~M. McNee, J.~A. Konstan, and G.~Lausen, ``Improving
  recommendation lists through topic diversification,'' in \emph{\Proceedings
  \International \Conference on World Wide Web}, 2005, pp. 22--32.

\bibitem{Vee08}
E.~Vee, U.~Srivastava, J.~Shanmugasundaram, P.~Bhat, and S.~A. Yahia,
  ``Efficient computation of diverse query results,'' in \emph{\Proceedings
  IEEE \International \Conference on Data Engineering}, 2008, pp. 228--236.

\bibitem{Yu09}
C.~Yu, L.~Lakshmanan, and S.~Amer-Yahia, ``It takes variety to make a world:
  diversification in recommender systems,'' in \emph{\Proceedings
  \International \Conference on Extending Database Technology: Advances in
  Database Technology}, 2009, pp. 368--378.

\bibitem{Vargas11}
S.~Vargas and P.~Castells, ``Rank and relevance in novelty and diversity
  metrics for recommender systems,'' in \emph{\Proceedings ACM \Conference on
  Recommender Systems}, 2011, pp. 109--116.

\bibitem{Drosou12}
M.~Drosou and E.~Pitoura, ``Dis{C} diversity: Result diversification based on
  dissimilarity and coverage,'' in \emph{\Proceedings VLDB Endowment}, vol.~6,
  no.~1, 2012, pp. 13--24.

\bibitem{Drosou17}
M.~Drosou, H.~Jagadish, E.~Pitoura, and J.~Stoyanovich, ``Diversity in big
  data: A review,'' \emph{Big data}, vol.~5, no.~2, pp. 73--84, 2017.

\bibitem{core-set-coverage02014}
P.~Indyk, S.~Mahabadi, M.~Mahdian, and V.~S. Mirrokni, ``Composable core-sets
  for diversity and coverage maximization,'' in \emph{33rd ACM
  SIGMOD-SIGACT-SIGART \Symposium on Principles of Database Systems}, 2014, pp.
  100--108.

\bibitem{Borodin12}
A.~Borodin, H.~Lee, and Y.~Ye, ``Max-{S}um diversification, monotone submodular
  functions and dynamic updates,'' \emph{\Proceedings ACM SIGACT-SIGMOD-SIGART
  \Symposium on Principles of Database Systems}, 2012.

\bibitem{Erkut94}
E.~Erkut, Y.~\"{U}lk\"{u}sal, and O.~Yenicerio\u{g}lu, ``A comparison of
  $p$-dispersion heuristics,'' \emph{Computers \& Operations Research},
  vol.~21, no.~10, pp. 1103--1113, 1994.

\bibitem{Lindley61}
D.~Lindley, ``Dynamic programming and decision theory,'' in \emph{Applied
  Statistics}, vol. 101, 1961, pp. 39--51.

\bibitem{Dynkin63}
E.~Dynkin, ``The optimum choice of the instant for stopping a {M}arkov
  process,'' in \emph{Sov. Math. Dokl}, 1963.

\bibitem{Bearden06}
J.~Bearden, ``A new secretary problem with rank-based selection and cardinal
  payoffs,'' in \emph{\Journal of Mathematical Psychology}, vol.~50, 2006, pp.
  58--59.

\bibitem{Babaioff07}
M.~Babaioff, N.~Immorlica, D.~Kempe, and R.~Kleinberg, ``A knapsack secretary
  problem with applications,'' in \emph{APPROX-RANDOM}, 2007.

\bibitem{Albers19}
S.~Albers and L.~Ladewig, ``New results for the $k$-secretary problem,'' in
  \emph{\International \Symposium on Algorithms and Computation}, ser. Leibniz
  \International \Proceedings in Informatics, vol. 149, 2019, pp. 18:1--18:19.

\bibitem{Mla22}
A.~Mladenovic, A.~J. Bose, H.~Berard, W.~L. Hamilton, S.~Lacoste-Julien,
  P.~Vincent, and G.~Gidel, ``Online adversarial attacks,'' \emph{The
  International Conference on Learning Representations}, 2022.

\bibitem{Broder09}
A.~Broder, A.~Kirsch, R.~Kumar, M.~Mitzenmacher, E.~Upfal, and
  S.~Vassilvitskii, ``The hiring problem and lake wobegon strategies,'' in
  \emph{SIAM \Journal on Computing}, vol.~39, 2009, pp. 1223--1255.

\bibitem{Eco1}
S.~K. Meerdink, S.~J. Hook, D.~A. Roberts, and E.~A. Abbott, ``The ecostress
  spectral library version 1.0,'' \emph{Remote Sensing of Environment}, vol.
  230, pp. 1--8, 2019.

\bibitem{Aster2}
A.~M. Baldridge, S.~J. Hook, C.~I. Grove, and G.~Rivera, ``The {ASTER} spectral
  library version 2.0,'' \emph{Remote Sensing of Environment}, vol. 113, pp.
  711--715, 2009.

\end{thebibliography}

\end{document}